\documentclass[a4paper,11pt]{article}
\pdfoutput=1 

\usepackage{jheppub} 

\usepackage[T1]{fontenc} 

\usepackage{pdfsync}
\usepackage{mathptmx}
\usepackage[utf8]{inputenc}
\usepackage[T1]{fontenc}
\usepackage{slashed}
\usepackage{times}
\usepackage{amssymb,amsfonts,amsmath,amsthm}
\usepackage{dsfont,bbm}
\usepackage{dcolumn}
\usepackage{epsf}
\usepackage{graphicx}
\usepackage[caption=false]{subfig}
\usepackage{dsfont}
\usepackage{bm}
\usepackage{slashed}
\usepackage[active]{srcltx}
\usepackage[usenames]{color}

\title{\boldmath Parton distributions: Functional complexity and Lorentz parametrization}

\author[a]{I.~V.~Anikin,}
\author[a]{L.~Szymanowski}

\affiliation[a]{National Centre for Nuclear Research (NCBJ),\\02-093 Warsaw, Poland}

\emailAdd{Igor.Anikin@ncbj.gov.pl}
\emailAdd{Lech.Szymanowski@ncbj.gov.pl}

\abstract{
In the paper we focus on the 
study of the functional complexity of the Lorentz parametrizing functions
in connection with the time-reversal transformations. 
We argue that the interactions encoded in the corresponding correlators
of non-local quark(-gluon) operators generate additional sources of functional complexity 
for the parametrizing functions which are not discussed in the literature. 
We also revisit the Lorentz parametrization of different correlators
given by the hadron matrix elements of the non-local operators.
The evidences for the new parametrizing function existence have been presented.}

\begin{document} 
\maketitle
\flushbottom

\section{Introduction}
\label{Intro}

The production in nucleon-nucleon collisions in the Drell-Yan (DY) processes
and the (semi-inclusive) deep inelastic scattering (SIDIS/DIS) processes provide still much information on the 
composite structure of hadrons. 
From the theoretical viewpoint, these processes give additional possibilities to  
study the new sorts of parton distribution functions which accumulate information on 
the transverse motion of quarks inside hadrons.
In fact, every of parton distribution functions is actually the Lorentz parametrizing 
function related to the given correlator. 
In this connection, it cannot be overestimated that 
the different properties of parametrizing functions which originate from the 
fundamental (discrete) symmetries play the very important role in investigations. 

In the most general case, the functions which depend on the parton momentum are given by
the hadron-hadron matrix element of non-local quark operators projected on the given $\Gamma$-combination.
Before factorization, it reads
\begin{eqnarray}
\label{Phi-1}
\Phi^{[\Gamma]}_{(\pm)}(k) = 
\int (d^4 z) e^{+i kz} \langle P,S | \bar\psi(0) \, \Gamma\,  
[0\,;z ]^{(\pm)}_A \, \psi(z)  | P,S \rangle^H,
\end{eqnarray}
where $[0\,;z ]^{(\pm)}_A$ stands for the future- and past-pointed Wilson line (WL)
\footnote{Throughout the paper, we use the standard notations for the plus and minus light-cone directions.}.
As usual, the $\Gamma$-com\-bi\-na\-tion corresponds to the Dirac $\gamma$-matrices which form the basis given by 
$\{ \mathbb{I},\, \gamma_5,\, \gamma^\mu,\, \gamma^\mu\gamma_5,\, \sigma^{\mu\nu} \}$.
The decomposition of $\Phi^{[\Gamma]}_{(\pm)}(k)$ based on the Lorentz covariance gives a number of parametrizing functions which, 
 in its turn, can be associated with the given distribution functions
(see, for example, 
\cite{Boglione:1999pz, Bacchetta:2004jz, Goeke:2005hb, Collins:2005rq, Anselmino:2008sga, Bastami:2018xqd}).
It is important to emphasize that the distribution functions as functions of the parton momentum fractions appear only after 
the factorization procedure has been applied. In the simplest $k_\perp$-independent case, we have 
\begin{eqnarray}
\label{Phi-1-facth}
\Phi^{[\Gamma]}_{(\pm)}(x) = \int (d^4 k) \delta(x - k^+/P^+) 
\Phi^{[\Gamma]}_{(\pm)}(k), \quad k=(k^+, k^-, \vec{\bf k}_\perp).
\end{eqnarray}
The necessity to study the $k_\perp$-dependence leads to the $k_\perp$-unintegrated functions $\Phi^{[\Gamma]}_{(\pm)}(x, k_\perp)$
and it requires, in a sense, the corresponding modification of factorization.   
 
In Eqn.~(\ref{Phi-1}),
$H$ indicates the Heisenberg representation (H-representation). 
In other words, we have to assume that all states and operators of Eqn.~(\ref{Phi-1}), roughly speaking, are ``dressed'' ones due to
the interactions. This is the well-known fact which is, however, forgotten or hidden in the literature quite frequently.
Meanwhile, the presence of interactions in the correlators provides not only the evolution of given distribution functions,
but it ensures the important fundamental properties of them.

In the present paper, we revisit the properties of distribution functions which stem from the time-reversal transformations together 
with the hermitian conjugations.  
Namely, we argue that the interactions encoded in the bound $in$- and/or $out$-states and in H-representation of 
quark(-gluon) operators forming the corresponding correlators result in additional sources of the functional complexity 
for the parametrizing functions. 
Then, based on the time-reversal and hermitian transforms, we demonstrate that 
the $k_\perp$-dependent  
parametrizing functions do not possess the certain symmetry properties under $k_\perp \to - k_\perp$.
This our finding is not stemmed from the past-pointed and future-pointed Wilson lines discussed in \cite{Boer:1997nt}.
At the same time, the usual time-reversal properties of functions can be restored but 
after the $k_\perp$-integration of the corresponding functions. 

In addition, we show that the careful taking into account of interactions in the correlators leads
to a new type of parametrizing functions which are associated with the inner quark structure defined by the quark spin.
These functions can contribute to the corresponding single quark spin asymmetry inside the unpolarized 
hadron or to the alignment $k_\perp$-dependent functions. 

\section{The time-reversal transforms of parameterizing functions }
\label{T-prop}

In this section, we dwell on the comprehensive analysis of the fundamental properties 
which have been imposed on the transverse momentum dependent distribution functions.
Namely, we focus on the time-reversal transforms of the correlators with and without 
the taking into account of interactions.

\subsection{The role of interactions in the correlators}
\label{Corr-H-repres}

At the beginning, we remind the role of H-representation (and the related interaction representation (I-representation))
in the definition of parametrizing functions.
As an example, let us consider the most typical (forward) Compton scattering (CS) amplitude which takes the form of
\begin{eqnarray}
\label{Amp-1}
{\cal A}_{\mu\nu} = \langle P| a^-_\nu(q) \, \mathbb{S}[\bar\psi, \psi, A] \, a^+_\mu(q) | P\rangle,
\end{eqnarray}
where $\mathbb{S}$-matrix is given by
\begin{eqnarray}
\mathbb{S}[\psi,\bar\psi, A]=T\text{exp}\Big\{ 
i \int (d^4 z) \big[ {\cal L}_{QCD}(z) + {\cal L}_{QED}(z)\big] \Big\}.
\nonumber
\end{eqnarray}
In Eqn.~(\ref{Amp-1}), in contrast to the photon Fock states the hadron states cannot be 
expressed through the relevant operators of creation and annihilation
\footnote{ The creation and annihilation hadron operators can be introduced with the help of the effective Lagrangian 
describing the transition of partons onto hadrons. 
This is the so-called effective quark-hadron Lagrangian of interaction \cite{Efimov:1993zg}.}.

Making used the commutation relations of creation (or annihilation) operators with $\mathbb{S}$-matrix 
(see, for example, Eqn.~(38.20b) of \cite{BogoShir}), 
\begin{eqnarray}
\label{Com-1}
&&\big[ a^\pm_\mu(q), \, \mathbb{S}[\bar\psi, \psi, A] \big]= \int (d^4 \xi) e^{\pm i q \xi} 
\frac{\delta \mathbb{S}[\bar\psi, \psi, A] }{ \delta A^\mu(\xi)} \quad \text{with}
\\
&&
\label{Der-S}
\frac{\delta \mathbb{S}[\bar\psi, \psi, A] }{ \delta A^\mu(\xi)} = 
T\Big\{  \int (d^4 z) \frac{\delta {\cal L}_{QED}(z)}{\delta A^\mu(\xi)} \, \mathbb{S}[\bar\psi, \psi, A] \Big\},
\end{eqnarray}
the CS-amplitude can be rewritten as 
\begin{eqnarray}
\label{Amp-2}
&&{\cal A}_{\mu\nu} = 
\int (d^4 \xi_1) (d^4 \xi_2) e^{- i q (\xi_1 - \xi_2)} 
\langle P| \frac{\delta^2 \mathbb{S}[\bar\psi, \psi, A] }{ \delta A^\mu(\xi_1) \delta A^\nu(\xi_2)} | P\rangle
\nonumber\\
&&
\Rightarrow \int (d^4 z) e^{- i q z} \langle P| T 
\Big\{ [\bar\psi(0)\gamma_\nu \psi(0)] \, 
[\bar\psi(z)\gamma_\mu \psi(z)]\,
\mathbb{S}[\bar\psi, \psi, A]  \Big\}
| P\rangle.
\end{eqnarray}
Using Wick's theorem and calculating the only quark operator contraction in Eqn.~(\ref{Amp-2}), 
we can readily obtain the simplest ``hand-bag'' diagram contribution to the CS-amplitude. 
Within the momentum representation, it reads 
 \begin{eqnarray}
\label{Amp-3}
&&{\cal A}_{\mu\nu}^{\text{hand-bag}} = 
\int (d^4 k) \,\text{tr} \big[ E_{\mu\nu}(k) \Phi(k) \big],
\end{eqnarray}
where 
\begin{eqnarray}
\label{E}
&&E_{\mu\nu}(k) = \gamma_\mu S(k+q) \gamma_\nu + \gamma_\nu S(k-q) \gamma_\mu,
\\
&&
\label{phifun}
\Phi(k)= \int (d^4 z) \, e^{ikz} 
\langle P| T \bar\psi(0) \psi(z) \mathbb{S}[\bar\psi, \psi, A] | P \rangle_c.
\end{eqnarray}
Here, the subscript $c$ denotes the connected diagram contributions which we only consider.

It is worth to notice that in Eqn.~(\ref{phifun}) the nonperturbative correlator has been written in the interaction representation.
It is more compact to use, however, the Heisenberg representation of this correlator, {\it i.e.}
\begin{eqnarray}
\label{phifun-H}
\Phi(k)= \int (d^4 z) \, e^{ikz} 
\langle P| \bar\psi(0) \psi(z) | P \rangle^H.
\end{eqnarray} 
We stress that the neglecting $H$ in Eqn.~(\ref{phifun-H}) may result in the wrong impression about the absence of interaction 
in the correlator.  

In the relevant correlators of Eqns.~(\ref{phifun}) and (\ref{phifun-H}), the interactions which are described by the 
$\mathbb{S}$-matrix generate, first, the Wilson lines ensuring the gauge invariance of non-local operators 
and, second, the contributions which are not being exponentiated. The latter can be associated with the 
evolutions of the corresponding functions or/and with the tensor structure of Lorentz parametrization, 
see section~\ref{Lor-Par-Int}. Since the Wilson lines can be eliminated  by the contour gauge use 
\cite{Anikin:2021osx,Anikin:2010wz,Anikin:2016bor}, 
in what follows we mainly focus on the non-exponentiated contributions of $\mathbb{S}$-matrix.  
  
The CS-amplitude of Eqn.~(\ref{Amp-3}) is not factorized yet. We omit the full description of the factorization procedure 
which is presented in the literature in detail. We would like to notice
that even after the applied factorization procedure the CS-amplitude contains the corresponding correlator 
(forming the soft part of amplitude)
where $\mathbb{S}$-matrix has to be included either in I-representation or in H-representation.

\subsection{On $in$- and $out$-states in the correlators}
\label{In-out-repres}

In this subsection, we remind the classical definitions of $in$- and $out$-states which appear in any correlators, 
see for example Eqns.~(\ref{phifun}) and (\ref{phifun-H}). The properties of these states play an important role for our 
study.

As a example, we consider
the physical (scalar hadron) states which are being described by the  $in$- and $out$-fields.
The  $in$- and $out$-fields are nothing but the asymptotically free fields. 
From the viewpoint of the standard classical scattering theory,
they can be treated as the asymptotes to the given trajectory. 
In other words, in order to find the correspondence between the 
initial  and final asymptotic states, 
we have to know the potential of interaction and to solve the corresponding equations of motion
(see \cite{Gasiorowicz:1966xra, Zavyalov:1990kv} for further details). 
Indeed, let us consider the following matrix element 
\begin{eqnarray}
\label{n-in-out-1}
\langle p_1, ..., p_n ; out | q_1, ..., q_m ; in \rangle = 
\langle  0 | a^-_{out}(p_1) ... a^-_{out}(p_n)  | q_1, ..., q_m ; in \rangle,
\end{eqnarray}
where the states have been constructed from the scalar hadron fields ($\varphi$-fields) and $\langle  0 |$ is a mathematical vacuum state.

Using the relation \cite{Gasiorowicz:1966xra, Zavyalov:1990kv}
\begin{eqnarray}
\label{n-in-out-2}
a^\pm_{out}(p_i) = \mathbb{S}^\dagger[\varphi] \otimes a^\pm_{in}(p_i) \otimes \mathbb{S}[\varphi] 
\end{eqnarray}
with the unitary $\mathbb{S}$-matrix describing the interaction of scalar fields $\varphi$,
Eqn.~(\ref{n-in-out-1}) takes the form of 
\begin{eqnarray}
\label{n-in-out-1-2}
&&\langle  0 | a^-_{out}(p_1) ... a^-_{out}(p_n)  | q_1, ..., q_m ; in \rangle=
\nonumber\\
&&
\langle  0 | a^-_{in}(p_1) ... a^-_{in}(p_n) \,\mathbb{S}[\varphi]\,  | q_1, ..., q_m ; in \rangle=
\langle p_1, ..., p_n ; in | \,\mathbb{S}[\varphi]\, | q_1, ..., q_m ; in \rangle.
\end{eqnarray}
Hence, we derive the well-known relation between $out$- and $in$-states which correspond to the final Fock state given by
\begin{eqnarray}
\label{n-in-out-1-3}
\langle p_1, ..., p_n ; out |=
\langle p_1, ..., p_n ; in | \, \mathbb{S}[\varphi].
\end{eqnarray}
It is important to notice that one-particle states meet the trivial condition
\begin{eqnarray}
\label{in-out-1part}
\langle  0 | a^-_{out}(p) = \langle  0 | a^-_{in}(p).
\end{eqnarray}
Indeed, the simple algebra gives us the following line of reasoning
\begin{eqnarray}
\label{in-out-1part-2}
\langle  0 | a^-_{out}(p) = \langle  0 | a^-_{in}(p) \, \mathbb{S}[\varphi] =
 \langle  0 | \left[ a^-_{in}(p), \, \mathbb{S}[\varphi] \right] + 
\langle  0 | \mathbb{S}[\varphi]\, a^-_{in}(p) = \langle  0 | a^-_{in}(p).
\end{eqnarray}
In this line of reasoning, we take into account that \cite{BogoShir}
\begin{eqnarray}
\label{in-out-1part-3}
\langle  0 | \left[ a^-_{in}(p), \, \mathbb{S}[\varphi] \right] =
\int (d^4\xi) e^{+ip\xi} \langle  0 | \frac{\delta \mathbb{S}[\varphi]}{\delta \varphi(\xi)} \sim \delta^{(4)}(p)
\end{eqnarray}
owing to the fact that the variation derivative of $\mathbb{S}[\varphi]$ gives only 
the tadpole contributions multiplying by the delta-function with the momentum conservation as an argument. 
And, if the hadron momentum $p$ is nonzero, the commutator term in Eqn.~(\ref{in-out-1part-2})
does not contribute.

Within the center-mass system, Eqn.~(\ref{n-in-out-1-3}) can be simplified for the case of two-particle states
with the help of partial wave expansion. It reads \cite{Gasiorowicz:1966xra}
 \begin{eqnarray}
\label{in-out-2part}
\langle p_1, p_2 ; out |=
\langle p_1, p_2 ; in | \, \mathbb{S}[\varphi] = \langle p_1, p_2 ; in | e^{i\delta_{p_1 p_2}(W)},
\end{eqnarray}
where $\delta_{p_1 p_2}(W)$ implies the scattering phase as a function of energy $W$.
Eqn.~(\ref{in-out-2part}) shows that even in the most simple case the $\mathbb{S}$-matrix indeed generates the complexity.

To conclude this subsection, we discuss the most general form of correlators given by 
\begin{eqnarray}
\label{corr-in-out-1}
\langle  P ; out | \widetilde{\cal{O}}(\bar\psi, \psi, A) | P ; in \rangle, 
\end{eqnarray}
where the quark-gluon operator $\widetilde{\cal{O}}(\bar\psi, \psi, A)$ 
is defined as ${\cal O} (\bar\psi, \psi, A)\, \mathbb{S}[\bar\psi, \psi, A]$ (see Eqn.~(\ref{phifun})).
Here, the hadron states of $\langle  P ; out |$ and $| P ; in \rangle$ are the one-particle states unless the composite parton (quark-gluon) 
structure of hadron has been incorporated. In the case of bound  hadron states, 
the annihilation operator of hadron should be replaced by the unknown, from the point of view of QCD, 
function $\mathsf A^-$ of annihilation operators of partons, {\it i.e.} we have 
\begin{eqnarray}
\label{Qg-st-1}
a^-_{out}(P) \Rightarrow \mathsf{A}^-_{out}\left( a^-_{out}(k_1), ..., a^-_{out}(k_n) | P= k_1+ ... + k_n \right),
\end{eqnarray}
and the similar expression can be written for the $in$-state.
The function $\mathsf{A}^-$ should meet the condition as
\begin{eqnarray}
\label{Cond-BS-1}
\langle 0 | \mathsf{A}^-_{out}\left( a^-_{out}(k_1), ..., a^-_{out}(k_n) \right) = 
\langle 0 | \mathsf{A}^-_{in}\left( a^-_{in}(k_1), ..., a^-_{in}(k_n) \right)
\end{eqnarray}
because in the limit where the hadron is described as a point-like particle 
one should get the condition of Eqn.~(\ref{in-out-1part}). 
Moreover, Eqn.~(\ref{Cond-BS-1}) resembles formally the condition known from the theory of 
group representations. Indeed, the necessary condition to construct the 
representation of a group is $F^\prime(x^\prime)\equiv U_g F(O_g x) = F(x)$ 
or $F(O_g x) = U^{-1}_g F(x)$ where $U_g$ is an operator that acts on the representation space and 
$O_g$ is an operator defined on the given group.

Hence, for the quark-gluon partons, we have the following relation
\begin{eqnarray}
\label{Qg-st-2-0}
&&
\mathcal{S}[\bar\psi, \psi, A] \otimes
\mathsf{A}^-_{in}\Big( \mathbb{S}^\dagger[\bar\psi, \psi, A] \otimes
a^-_{in}(k_1) \otimes \mathbb{S}[\bar\psi, \psi, A] , ... \Big)
\otimes \mathcal{S}^\dagger[\bar\psi, \psi, A]
= 
\nonumber\\
&& 
 \mathsf{A}^-_{in} \left( a^-_{in}(k_1), ..., a^-_{in}(k_n) \right) ,
\end{eqnarray}
or 
\begin{eqnarray}
\label{Qg-st-2}
&&\langle 0 | \mathsf{A}^-_{in}\left( a^-_{out}(k_1), ..., a^-_{out}(k_n) | P= k_1+ ... + k_n \right)= 
\nonumber\\
&& 
\langle 0 | \mathcal{S}^\dagger[\bar\psi, \psi, A] \otimes \mathsf{A}^-_{in} \left( a^-_{in}(k_1), ..., a^-_{in}(k_n) | P= k_1+ ... + k_n \right) 
\otimes \mathcal{S}[\bar\psi, \psi, A],
\end{eqnarray}
where 
$\mathcal{S}[\bar\psi, \psi, A]$ describes effectively the quark-gluon interactions inside the hadron.
In other words, the relation given by Eqn.~(\ref{Qg-st-2}) can be treated as a definition of the unknown 
$\mathcal{S}$-operator which reflects the fact that the bound state is being always considered 
as the {\it dressed} (by the quark-gluon interaction) states. That is, we adhere the conception 
according to which 
the {\it physical} bound states have been treated as the states which cannot be described by the free Lagrangian. 
The physical statement on that the bound states are always dressed is equivalent to the 
mathematical requirement on the nullification of the hadron renormalization constant, $\mathbb{Z}_h=0$.   
This is the so-called compositeness condition, see  
\cite{Efimov:1993zg} and Appendix~\ref{App1} for the clarifying details.
Notice that for our further study the explicit form of $\mathcal{S}$-operator is not required, 
however the presence of this operator supports additionally the statement on the complexity of the 
given correlator.

With these, Eqn.~(\ref{corr-in-out-1}) can be presented as 
\begin{eqnarray}
\label{corr-in-out-2}
&&\langle 0| \mathsf{A}^-_{in}\left( a^-_{out}(k_1), ..., a^-_{out}(k_n) | P= k_1+ ... + k_n \right)  \widetilde{\cal{O}}(\bar\psi, \psi, A) | P ; in \rangle=
\nonumber\\
&&
\langle 0| \mathsf{A}^-_{in}\left( a^-_{in}(k_1), ..., a^-_{in}(k_n) | P= k_1+ ... + k_n \right)  
\mathcal{S}[\bar\psi, \psi, A] \, 
\widetilde{\cal{O}}(\bar\psi, \psi, A) | P ; in \rangle=
\nonumber\\
&&
\langle P ; in | \mathcal{S}[\bar\psi, \psi, A] \, 
\widetilde{\cal{O}}(\bar\psi, \psi, A) | P ; in \rangle.
\end{eqnarray}
The principle conclusion from this consideration is that the trivial relation between one-hadron $in$-
and $out$-state given by Eqn.~(\ref{in-out-1part}) has been non-trivially modified if the 
quark-gluon interactions are included inside the given hadron.  

\subsection{The time-reversal properties without interactions in correlators}
\label{No-Int-Corr}

We are going over to the discussion of properties which are originated from the time-reversal transformations.
The time-reversal transforms imply that 
\begin{eqnarray}
\label{T-rev-tr-1}
\langle \Phi_2 | U^+_T U_T \mathbb{O}(0,z) | \Phi_1 \rangle = 
\Big [  \langle \Phi_2 | \mathbb{O}(0,z) | \Phi_1 \rangle
\Big]^\dagger,
\end{eqnarray}
where $\mathbb{O}$ is an arbitrary nonlocal operator with the closed Dirac indices,
the operator $U_T$ is acting on the Fock states.
In Eqn.~(\ref{T-rev-tr-1}), the correlator does not contain any interactions, $\mathbb{S}[\bar\psi, \psi, A] = \mathbb{I}$.

For the sake of definiteness, we suppose that 
\begin{eqnarray}
\label{O-Oper}
\mathbb{O}(0,z)= \bar\psi(0) \gamma^+ \psi(z), \quad
| \Phi_1 \rangle = | P, S \rangle, \quad \langle \Phi_2 | = \langle P, S |.
\end{eqnarray}
As mentioned, the Wilson lines in $\mathbb{O}(0,z)$ have been eliminated by the 
contour gauge of axial type \cite{Anikin:2021osx,Anikin:2010wz,Anikin:2016bor} and, in the context of our study, 
they are not considered as the effect of the interaction presence.

Hence, Eqn.~(\ref{T-rev-tr-1}) takes the form of 
\begin{eqnarray}
\label{T-rev-tr-2}
\langle \tilde P, \tilde S |  \bar\psi(0) \gamma^- \psi(\tilde z)  | \tilde P, \tilde S \rangle = 
\Big[ 
\langle P, S |  \bar\psi(0) \gamma^+ \psi(z)  | P, S \rangle
\Big]^\dagger,
\end{eqnarray}
where 
\footnote{More precisely, if we use the standard definition 
of covariant and contravariant vectors, we have $\tilde z^\mu =(-z_0, \vec{\bf z})= - z_\mu$ and 
$\tilde A^\mu=(A_0, -\vec{\bf A})=A_\mu$. However, this is irrelevant for the subject of our discussion}
\begin{eqnarray}
\tilde z =(-z_0, \vec{\bf z}),\quad \tilde A=(A_0, -\vec{\bf A})\quad \text{for}\quad A=(P, S).
\end{eqnarray}

Concentrating on the {\it l.h.s.} of Eqn.~(\ref{T-rev-tr-2}) and extracting only $(b^+\, b^-)$-combination 
of quark creation operators, the Fourier transforms of quark operators give the following
representation
\begin{eqnarray}
\label{Par-1}
&&
\langle \tilde P, \tilde S |  \bar\psi(0) \gamma^- \psi(\tilde z)  | \tilde P, \tilde S \rangle = 
\int (d^4 \tilde k) \,e^{-i \tilde k \tilde z} \,\Phi^{[\gamma^-]}(\tilde k, \tilde P, \tilde S) 
\stackrel{\text{L. par.}}{\Longrightarrow}
\nonumber\\
&&
\int (d\tilde k^-)(d^2 \widetilde{\vec{\bf k}}_\perp) \,e^{-i \tilde k^- \tilde z^+ + i \widetilde{\vec{\bf k}}_\perp \widetilde{\vec{\bf z}}_\perp} \,
\Big\{ 
\tilde P^- f_1(\tilde k^-, \tilde k_\perp) + \epsilon^{- \tilde P \tilde k_\perp \tilde S} f^{\perp\, \text{(n.)}}_{1\, T} (\tilde k^-, \tilde k_\perp) + .....
\Big\},
\end{eqnarray}
where the normalized function $f^{\perp\, \text{(n.)}}_{1\, T} = f^\perp_{1\, T}/m_N$ has been introduced and 
the Lorentz parametrization (L. par.) has been applied. Also, in this representation we implement the integration over $d\tilde k^+$. 

Then, taking into account that 
\begin{eqnarray}
\label{mom-pl-mn}
&& \tilde k^\pm = k^\mp, \quad \tilde P^\pm = P^\mp, \quad  \tilde z^{\mp} = - z^\pm,
\nonumber\\
&&
\widetilde{\vec{\bf k}}_\perp = - \vec{\bf k}_\perp, \quad 
 \widetilde{\vec{\bf P}}_\perp =  - \vec{\bf P}_\perp ,
\quad \widetilde{\vec{\bf z}}_\perp = \vec{\bf z}_\perp
\end{eqnarray}
Eqn.~(\ref{Par-1}) can be rewritten as 
\begin{eqnarray}
\label{Par-2}
&&
\langle \tilde P, \tilde S |  \bar\psi(0) \gamma^- \psi(\tilde z)  | \tilde P, \tilde S \rangle 
\stackrel{\text{L. par.}}{\Longrightarrow}
\nonumber\\
&&
\int (d k^+)(d^2 \widetilde{\vec{\bf k}}_\perp) \,e^{ i k^+ z^- + i \widetilde{\vec{\bf k}}_\perp \widetilde{\vec{\bf z}}_\perp} \,
\Big\{ 
P^+ f_1(k^+, \tilde k_\perp) - \epsilon^{+ P \tilde k_\perp \tilde S} f^{\perp\, \text{(n.)}}_{1\, T} (k^+, \tilde k_\perp) + .....
\Big\}=
\nonumber\\
&&
\int (d k^+)(d^2 \vec{\bf k}_\perp) \,e^{ i k^+ z^- - i \vec{\bf k}_\perp \vec{\bf z}_\perp} \,
\Big\{ 
P^+ f_1(k^+, - k_\perp) - \epsilon^{+ P k_\perp S} f^{\perp\, \text{(n.)}}_{1\, T} (k^+, - k_\perp) + .....
\Big\},
\end{eqnarray}
where in the second term we use that  
\begin{eqnarray}
\epsilon^{- \tilde P \tilde k_\perp \tilde S} = \tilde P^- \epsilon^{- + \tilde k_\perp \tilde S} = 
P^+ \epsilon^{- + \tilde k_\perp \tilde S}= \epsilon^{ P + \tilde k_\perp \tilde S} = - \epsilon^{+ P \tilde k_\perp \tilde S}.
\end{eqnarray}
  
Let us now consider the {\it r.h.s.} of Eqn.~(\ref{T-rev-tr-2}) where one can see the following subtlety.
Namely, the Lorentz parametrization (or decomposition) can be performed both {\it (a)} before the Hermitian conjugation 
and {\it (b)} after this conjugation:
\begin{itemize}
\item
the first way {\it (a)} leads to 
\begin{eqnarray}
\label{Par-rhs-1}
&&
\Big[ 
\langle P, S |  \bar\psi(0) \gamma^+ \psi(z)  | P, S \rangle
\Big]^\dagger
\stackrel{\text{L. par.}}{\Longrightarrow}
\nonumber\\
&&
\Big[
\int (d k^+)(d^2 \vec{\bf k}_\perp) \,e^{ - i k^+ z^- + i \vec{\bf k}_\perp \vec{\bf z}_\perp} \,
\Big\{ 
P^+ f_1(k^+, k_\perp) + \epsilon^{+ P k_\perp S} f^{\perp\, \text{(n.)}}_{1\, T} (k^+, k_\perp) + .....
\Big\}
\Big]^\dagger=
\nonumber\\
&&
\int (d k^+)(d^2 \vec{\bf k}_\perp) \,e^{  i k^+ z^- - i \vec{\bf k}_\perp \vec{\bf z}_\perp} \,
\Big\{ 
P^+ [f_1(k^+, k_\perp)]^* + \epsilon^{+ P k_\perp S} [f^{\perp\, \text{(n.)}}_{1\, T} (k^+, k_\perp)]^* + .....
\Big\};
\end{eqnarray}
\item
while the second way {\it (b)} gives 
\begin{eqnarray}
\label{Par-rhs-2}
&&
\Big[ 
\langle P, S |  \bar\psi(0) \gamma^+ \psi(z)  | P, S \rangle
\Big]^\dagger=
\langle P, S |  \bar\psi(z) \gamma^+ \psi(0)  | P, S \rangle 
=
\langle P, S |  \bar\psi(0) \gamma^+ \psi(-z)  | P, S \rangle
\stackrel{\text{L. par.}}{\Longrightarrow}
\nonumber\\
&&
\int (d k^+)(d^2 \vec{\bf k}_\perp) \,e^{  i k^+ z^- - i \vec{\bf k}_\perp \vec{\bf z}_\perp} \,
\Big\{ 
P^+ f_1(k^+, k_\perp) + \epsilon^{+ P k_\perp S} f^{\perp\, \text{(n.)}}_{1\, T} (k^+, k_\perp) + .....
\Big\}.
\end{eqnarray}
\end{itemize}
Comparing Eqns.~(\ref{Par-rhs-1}) and (\ref{Par-rhs-2}), 
we derive the following properties
\begin{eqnarray}
\label{Prop-1}
[f_1(k^+, k_\perp)]^* = f_1(k^+, k_\perp), \quad [f^{\perp\, \text{(n.)}}_{1\, T} (k^+, k_\perp)]^*=f^{\perp\, \text{(n.)}}_{1\, T} (k^+, k_\perp)
\end{eqnarray}
provided (here, we do not specify the $in$- and $out$-states because 
the hadron states have been considered as one-particle states, see Eqn.~(\ref{in-out-1part}))
\begin{eqnarray}
\label{C-1}
\big[ | P, S \rangle \big]^\dagger = \langle P, S |.
\end{eqnarray}
These properties together with Eqns.~(\ref{T-rev-tr-2}) and (\ref{Par-2}) result in 
\begin{eqnarray}
\label{Prop-2}
&&
f_1(k^+, - k_\perp)=[f_1(k^+, k_\perp)]^* = f_1(k^+, k_\perp) \quad \rightarrow \quad \text{T-even} 
\nonumber\\
&&
- f^{\perp\, \text{(n.)}}_{1\, T} (k^+, - k_\perp)=[f^{\perp\, \text{(n.)}}_{1\, T} (k^+, k_\perp)]^*=f^{\perp\, \text{(n.)}}_{1\, T} (k^+, k_\perp)
\quad \rightarrow \quad \text{T-odd}.
\end{eqnarray}
The properties of Eqn.~(\ref{Prop-2}) are well-known in the literature for the transverse momentum dependent functions.

\subsection{The influence of interactions on the time-reversal properties}
\label{With-Int-Corr}

In the preceding subsection the interactions in correlators have been excluded. 
Now, we analyse the influence of interactions in the relevant correlators on
the properties of parametrizing functions with respect to $k_\perp \to - k_\perp$.

\begin{figure}[tbp]
\centering 
\includegraphics[width=.49\textwidth]{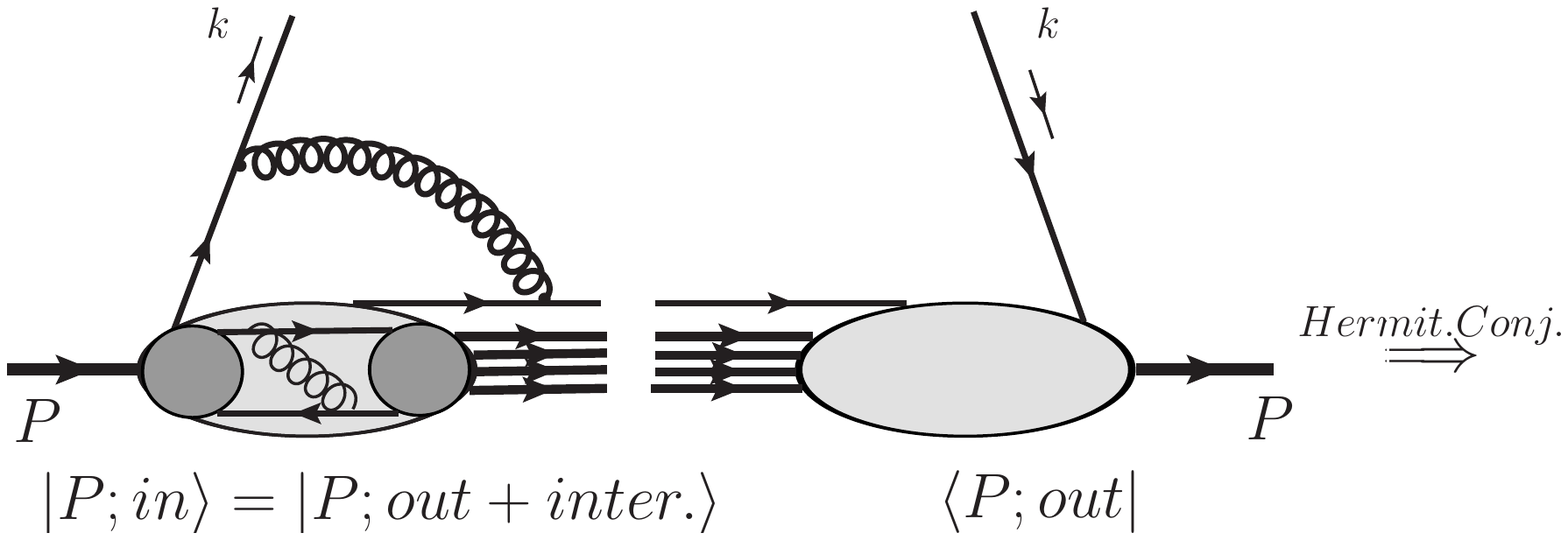}\,
\includegraphics[width=.49\textwidth]{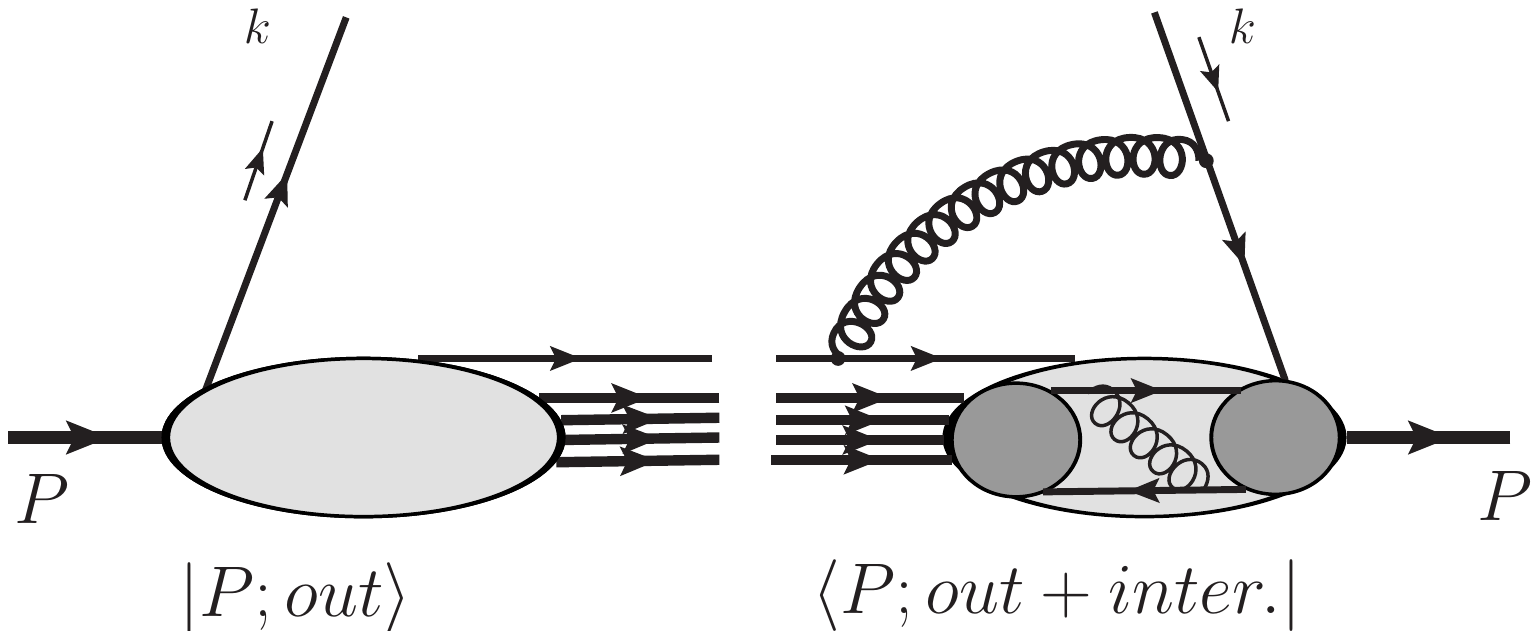}
\vspace{-7cm}
\caption{\label{Fig-S-1} The hermitian conjugation: the left panel - before conjugation; 
the right panel - after conjugation.}
\end{figure}

First, we dwell on the Hermitian conjugation which appears in the {\it r.h.s.} of Eqn.~(\ref{T-rev-tr-2}).
Taking into account $\mathbb{S}$-matrix, we have the following 
\begin{eqnarray}
\label{Par-rhs-S-1}
\Big[ 
\langle P, S |  \bar\psi(0) \gamma^+ \psi(z) \, \mathbb{S}[\bar\psi, \psi, A] | P, S \rangle
\Big]^\dagger.
\end{eqnarray}
For the sake of illustration, it is convenient to expand 
$\mathbb{S}$-matrix, say, upto the second order of interaction, see Fig.~\ref{Fig-S-1}, {\it i.e.}
\begin{eqnarray}
\label{Par-rhs-S-1-2}
\Big[ 
\langle P, S |  \bar\psi(0) \gamma^+ \psi(z) \, \mathbb{S}^{(2)}[\bar\psi, \psi, A] | P, S \rangle
\Big]^\dagger.
\nonumber
\end{eqnarray}
We stress that the order of expansion does not play any role for our final conclusions. 

Having used the results of subsection \ref{In-out-repres}, we have the following 
\begin{eqnarray}
\label{in-out-1}
&&\langle P, S | \equiv \langle P, S; out | = 
\langle P, S; in | \mathcal{S}[\bar\psi, \psi, A],
\nonumber\\
&&
| P, S \rangle \equiv | P, S; in \rangle = 
 \mathcal{S}^\dagger[\bar\psi, \psi, A]  | P, S; out \rangle.
\end{eqnarray}
In contract to Eqn.~(\ref{C-1}), the Hermitian conjugation of Eqn.~(\ref{in-out-1}) shows that (see Fig.~\ref{Fig-S-1})   
\begin{eqnarray}
\label{C-2}
\Big[ | P, S; in  \rangle \Big]^\dagger  = \langle P, S; out | \,  \mathcal{S}[\bar\psi, \psi, A] \not= 
\langle P, S; out | \quad \text{etc.}
\end{eqnarray}
owing to the included interactions of quarks and gluons inside the hadron.
And as such, the correlator of Eqn.~(\ref{Par-rhs-S-1}) can be presented in the form of 
\begin{eqnarray}
\label{Par-rhs-S-2}
&&\Big[ 
\langle P, S; out | \bar\psi(0) \gamma^+ \psi(z) \, \mathbb{S}[\bar\psi, \psi, A] 
\mathcal{S}^\dagger [\bar\psi, \psi, A]  | P, S; out \rangle
\Big]^\dagger=
\\
&&
\label{Par-rhs-S-2-2}
\langle P, S; out | \mathcal{S} [\bar\psi, \psi, A] \mathbb{S}^\dagger[\bar\psi, \psi, A]\, \bar\psi(z) \gamma^+ \psi(0) 
 | P, S; out \rangle.
\end{eqnarray}
From these, it is clear that the functions parametrizing the given correlator must be complex functions
thanks for the presence of the product $\mathbb{S}[\bar\psi, \psi, A] \mathcal{S}^\dagger [\bar\psi, \psi, A] $.
The difference between the parametrizing functions of Eqns.~(\ref{Par-rhs-S-2})  and (\ref{Par-rhs-S-2-2})
appears only in the imaginary parts, see below. 

Thus, the Lorentz parametrization is sensitive to the correlator which we deal with.
Namely, 
{\it the Hermitian conjugation is not a ``commutative'' operation with the Lorentz parametrization
if the interactions are presented}.
Hence, the parametrizing functions in Eqns.~(\ref{Par-rhs-1}) and (\ref{Par-rhs-2}) are not identical ones and 
the properties defined in Eqn.~(\ref{Prop-1}) are modified by 
\begin{eqnarray}
\label{Prop-1-2}
[f^{(a)}_1(k^+, k_\perp)]^* = f^{(b)}_1(k^+, k_\perp), 
\quad [f^{\perp\, \text{(n.)}\, (a)}_{1\, T} (k^+, k_\perp)]^*=f^{\perp\, \text{(n.)}\, (b)}_{1\, T} (k^+, k_\perp).
\end{eqnarray}
or 
\begin{eqnarray}
\label{Prop-1-3}
&&\Re\text{e}f^{(a)}_1(k^+, k_\perp) = \Re\text{e} f^{(b)}_1(k^+, k_\perp), \,
- \Im\text{m}f^{(a)}_1(k^+, k_\perp) = \Im\text{m} f^{(b)}_1(k^+, k_\perp);
\\
&&
\Re\text{e}f^{\perp\, \text{(n.)}\, (a)}_{1\, T} (k^+, k_\perp)=\Re\text{e}f^{\perp\, \text{(n.)}\, (b)}_{1\, T} (k^+, k_\perp),
\,
- \Im\text{m}f^{\perp\, \text{(n.)}\, (a)}_{1\, T} (k^+, k_\perp)=\Im\text{m}f^{\perp\, \text{(n.)}\, (b)}_{1\, T} (k^+, k_\perp).
\nonumber
\end{eqnarray}

So, the time-reversal transformations given by
\begin{eqnarray}
\label{T-tr-Int-1}
\langle \tilde P, \tilde S; in |  \bar\psi(0) \gamma^- \psi(\tilde z)  \, \mathbb{S}[\bar\psi, \psi, A]| \tilde P, \tilde S; out \rangle = 
\Big[ 
\langle P, S; out |  \bar\psi(0) \gamma^+ \psi(z) \, \mathbb{S}[\bar\psi, \psi, A]| | P, S; in  \rangle
\Big]^\dagger
\end{eqnarray} 
together with Eqn.~(\ref{Par-rhs-1}) 
result in the following properties for the corresponding parametrizing functions
\footnote{Here, the specification of $(a)$ and $(b)$ ways is irrelevant for the complex functions.}:
\begin{eqnarray}
\label{Prop-2-2}
&&f_1(k^+, - k_\perp)=[f_1(k^+, k_\perp)]^*,\quad 
- f^{\perp\, \text{(n.)}}_{1\, T} (k^+, - k_\perp)=[f^{\perp\, \text{(n.)}}_{1\, T} (k^+, k_\perp)]^*,
\nonumber\\
&&
\big\{ f_1(k^+, k_\perp), f^{\perp\, \text{(n.)}}_{1\, T} (k^+, - k_\perp) \big\} \in \mathbb{C}
\end{eqnarray}  
which demonstrate that there are no the definite properties under the replacement $k_\perp \to - k_\perp$ in 
contract to the standard consideration. 
Again, the usual time-reversal properties are restored by the $k_\perp$-integrations.

The functional complexity of parametrizing functions we have discovered can be manifested in the DY-like processes 
with the essential contributions from the gluon poles \cite{Anikin:2022ocg}.

%
%

\section{The Lorentz parametrization of correlators including the interactions}
\label{Lor-Par-Int}

In the preceding section, it has been shown that due to the different sources of interactions in the correlators, 
{\it i.e.} in the hadron matrix element of quark-gluon operators,
all of the Lorentz parametrizing functions become the complex functions.
Now, we study the influence of interactions on the Lorentz parametrization in 
order to find new possible parametrizing functions.
For this aim, the $in$- and $out$-states can be left without the specification explained in the preceding section
because we now focus on the tensor structure of relevant correlators.

\begin{figure}[tbp]
\centering 
\includegraphics[width=.49\textwidth]{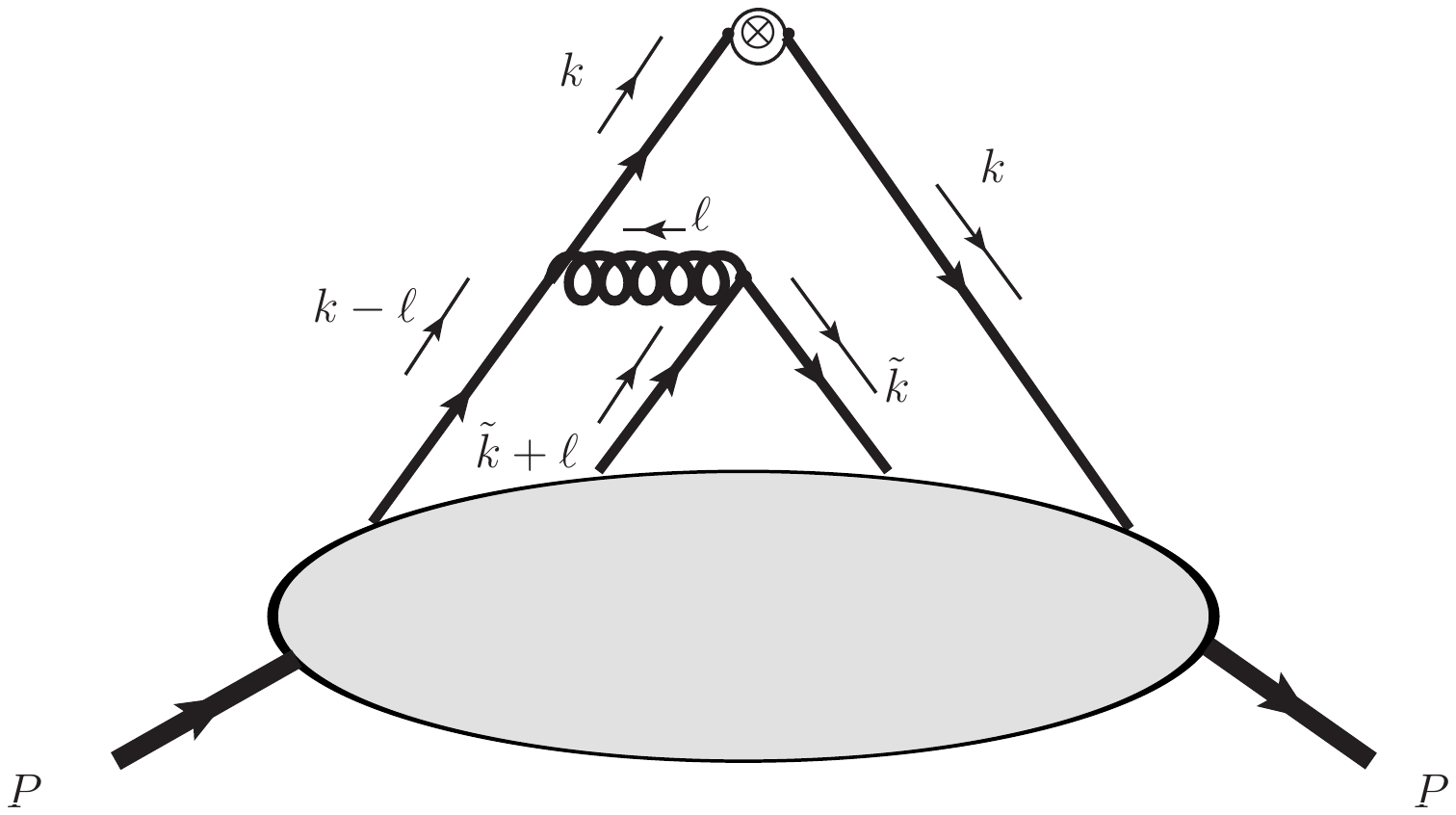}\,
\includegraphics[width=.49\textwidth]{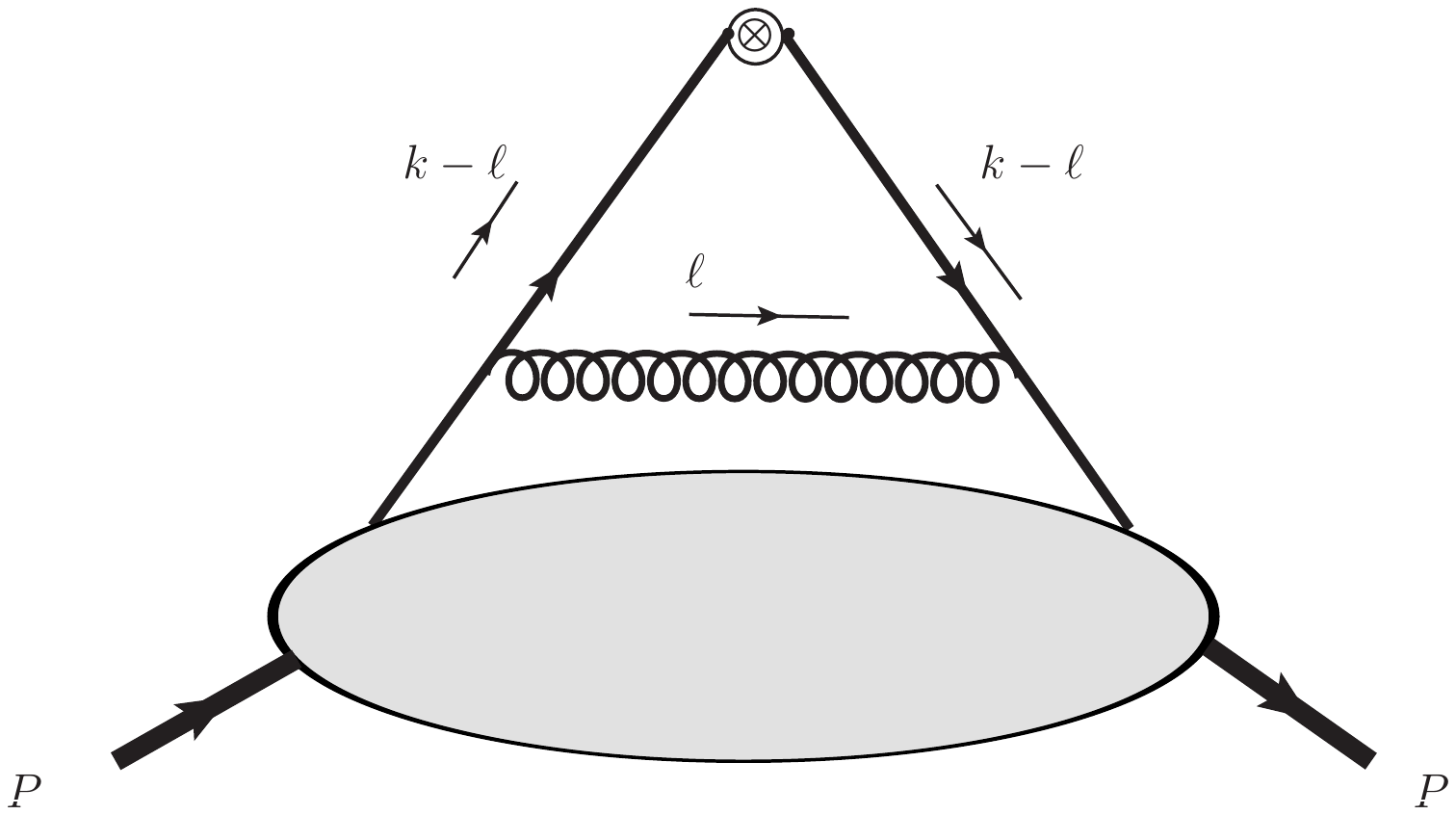}
\vspace{-6cm}
\caption{\label{Fig-S-1-2} The types of loop integrations in the corresponding correlators: 
the left panel corresponds to the demonstration of the implicit loop integrations defined the Lorentz structure; 
the right panel -- to the explicit lop integrations contributing to the evolution integration kernels.}
\end{figure}

We begin with the vector (the plus light-cone projection) correlator written in the interaction representation. It reads
\footnote{The symbol of time-ordering is omitted.}
\begin{eqnarray}
\label{Phi-1-2}
\Phi^{[\gamma^+]}(k) = 
\int (d^4 z) e^{+i kz} \langle P,S | \bar\psi(0) \, 
\gamma^+\, \psi(z) \, \mathbb{S}[\psi,\bar\psi, A] \,| P,S \rangle.
\end{eqnarray} 
Here, the $\mathbb{S}$-matrix generates the explicit and implicit loop integrations (modulo the Wilson lines
which are irrelevant within the contour gauge, see below).
The explicit loop integrations, see the right panel of Fig.~\ref{Fig-S-1-2}, are responsible for the 
forming of evolution integral kernels. While the implicit loop integrations, 
see the left panel of Fig.~\ref{Fig-S-1-2} as a particular example, are determining the Lorentz 
structures of the correlators.

 %
%
 
\begin{figure}[tbp]
\centering 
\includegraphics[width=.39\textwidth]{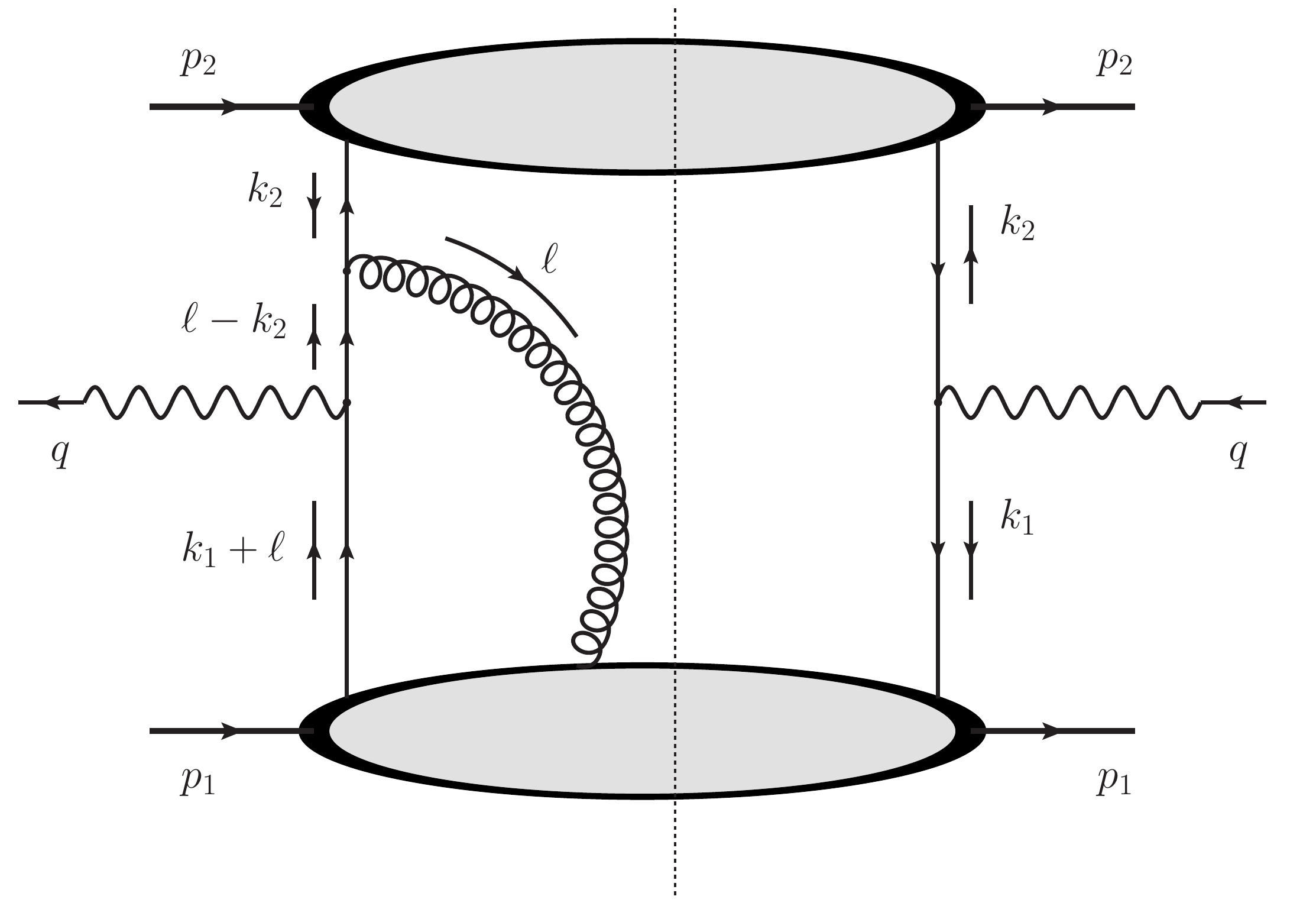}\,
\includegraphics[width=.39\textwidth]{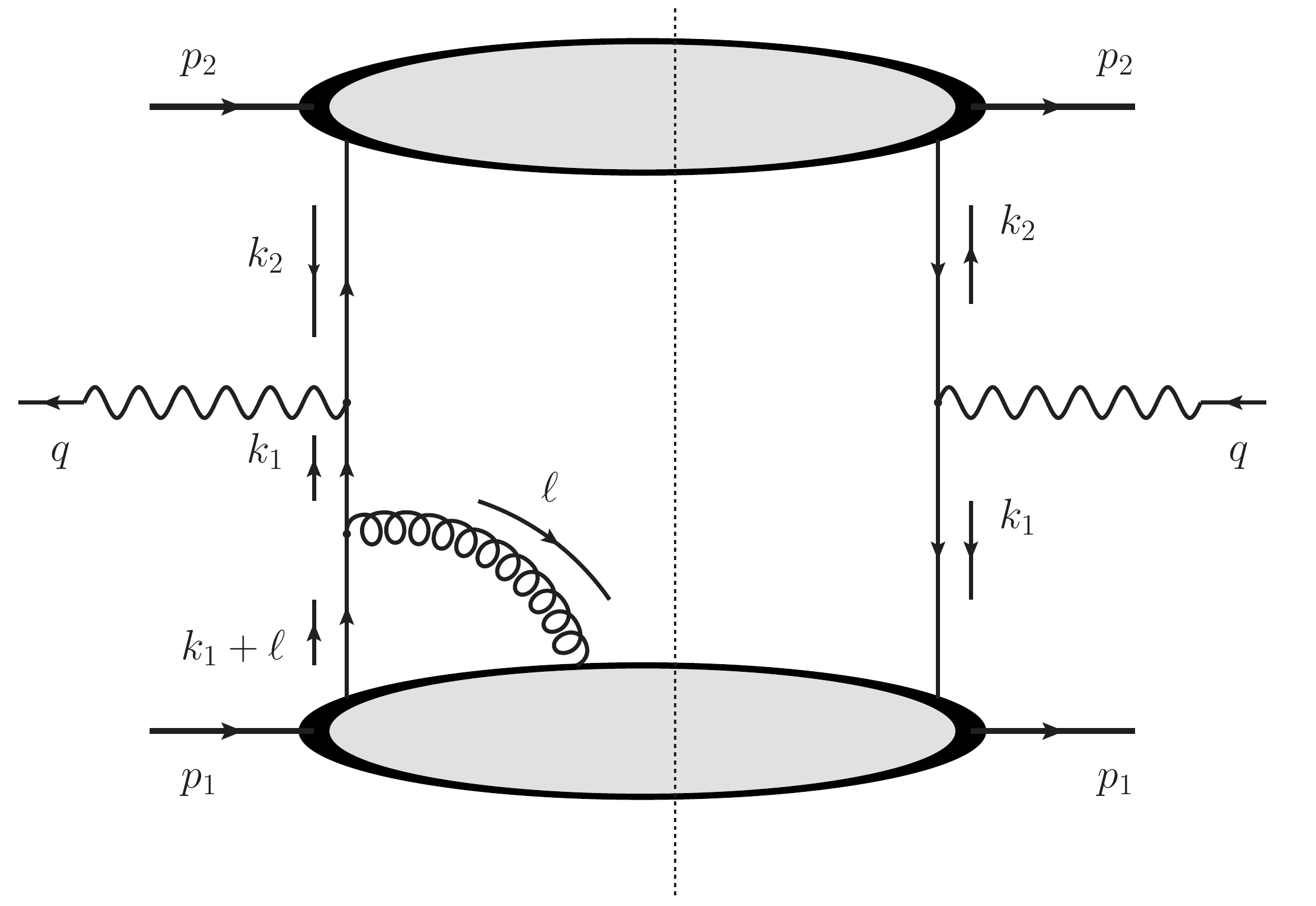}
\vspace{0.5cm}
\caption{\label{Fig-S-3} The standard diagram of DY-process, see 
the left panel, and the non-standard diagram of DY-process, see 
the right panel.}
\end{figure}

It is now a time to make the important comments on the Wilson lines.
In Eqn.~(\ref{Phi-1-2}), the  Wilson lines, which ensure the gauge invariance of non-local operators, 
are not shown because they can be eliminated by the corresponding contour gauge \cite{Anikin:2021osx,Anikin:2010wz,Anikin:2016bor}.
This point requires the additional explanations.
Following to \cite{Anikin:2010wz,Anikin:2016bor}, all the gluon radiation contributions that appears in the 
corresponding quark-gluon correlators of the Drell-Yan hadron tensor can be separated out in the three 
classes \footnote{The standard and non-standard diagrams have been defined in  \cite{Anikin:2010wz,Anikin:2016bor}}: 
\begin{itemize}
\item ({\it i}) the longitudinal $A^+$ and transverse $A^\perp_i$ ($i=1,2$) gluons from the {\it standard} diagram
(see the left panel of Fig.~\ref{Fig-S-3}) 
which are being exponentiated in the corresponding Wilson lines; 
\item ({\it ii}) the  longitudinal $A^-$ and transverse $A^\perp_i$ gluons from the {\it non-standard} diagram 
(see the right panel of Fig.~\ref{Fig-S-3}) 
which are being exponentiated in the other Wilson lines;  
\item ({\it iii}) the transverse $A^\perp_i$ gluons from both the standard and non-standard diagrams 
which {\it cannot} be exponentiated in the corresponding Wilson lines, but they construct the higher 
twist quark-gluon correlators.
\end{itemize}
%
%

Notice that, in contrast to the Wilson lines with longitudinal gluons, 
the nullification of the exponential functional in the relevant Wilson lines,
\begin{eqnarray}
\label{WL-exm}
W \left(A_i^\perp | P({\bf x}_0^\perp; {\bf x}^\perp)\right)=
\mathbb{P}\text{exp}
\Big\{ ig\int_{{\bf x}_0^\perp}^{{\bf x}^\perp} d \omega_\perp^i A^i_\perp(x^+_0, x^-_0, \omega_\perp)\Big\},
\end{eqnarray}
is related to the nullification of the full integral (but not the integrand) containing $A^\perp_i$, see  \cite{Anikin:2021osx}.  
In other words, in this case the trivialization of the Wilson lines does not lead to the absence of $A^\perp_i$.
Also, in the context of the contour gauge use, it is important to stress that the given contour gauge is not 
equivalent to the local axial gauge in the same manner as the given vector 
is not equivalent to its projection on the non-trivial direction \cite{Anikin:2021osx}. 
By definition, the trivial direction needed for the relevant projection coincides with the giving vector. 
Therefore, in discussing the interaction influence, the Wilson lines have been excluded from our consideration.
This our ``assumption'' does not effect on our principal conclusions.   

Among the standard parametrizing functions which are associated with the vector correlator 
(see, for example, \cite{Goeke:2005hb}),
we introduce the functions that are accompanying the Lorentz tensor with the 
quark spin (covariant) vector provided the interaction has been included in the correlator \cite{Anikin:2021zxl}. 
In particular, we have argued that 
the new type of functions is not excluded in the parametrization:  
\begin{eqnarray}
\label{Phi-plus-0}
&&\Phi^{[\gamma^+]}(k) = i \epsilon^{+ - V_\perp s_\perp} \tilde f_1^{(V)} (x; \,k^2_\perp) + ....
\end{eqnarray} 
where $s_\perp$ stands for the quark spin axial-vector and $V_\perp = \{P_\perp; k_\perp \}$.

In \cite{Anikin:2021zxl}, in order to prove the existence of this type of functions the second order of  
the $\mathbb{S}$-matrix decomposition over the strong coupling constant and the Fierz transformations applied 
for the relevant four-fermion combination have been used. 
Here, we give an alternative proof working with the fourth order of decomposition over the strong coupling constant and 
without the mentioned four-fermion Fierz transforms. 

%
%
\begin{figure}[tbp]
\centering 
\includegraphics[width=.49\textwidth]{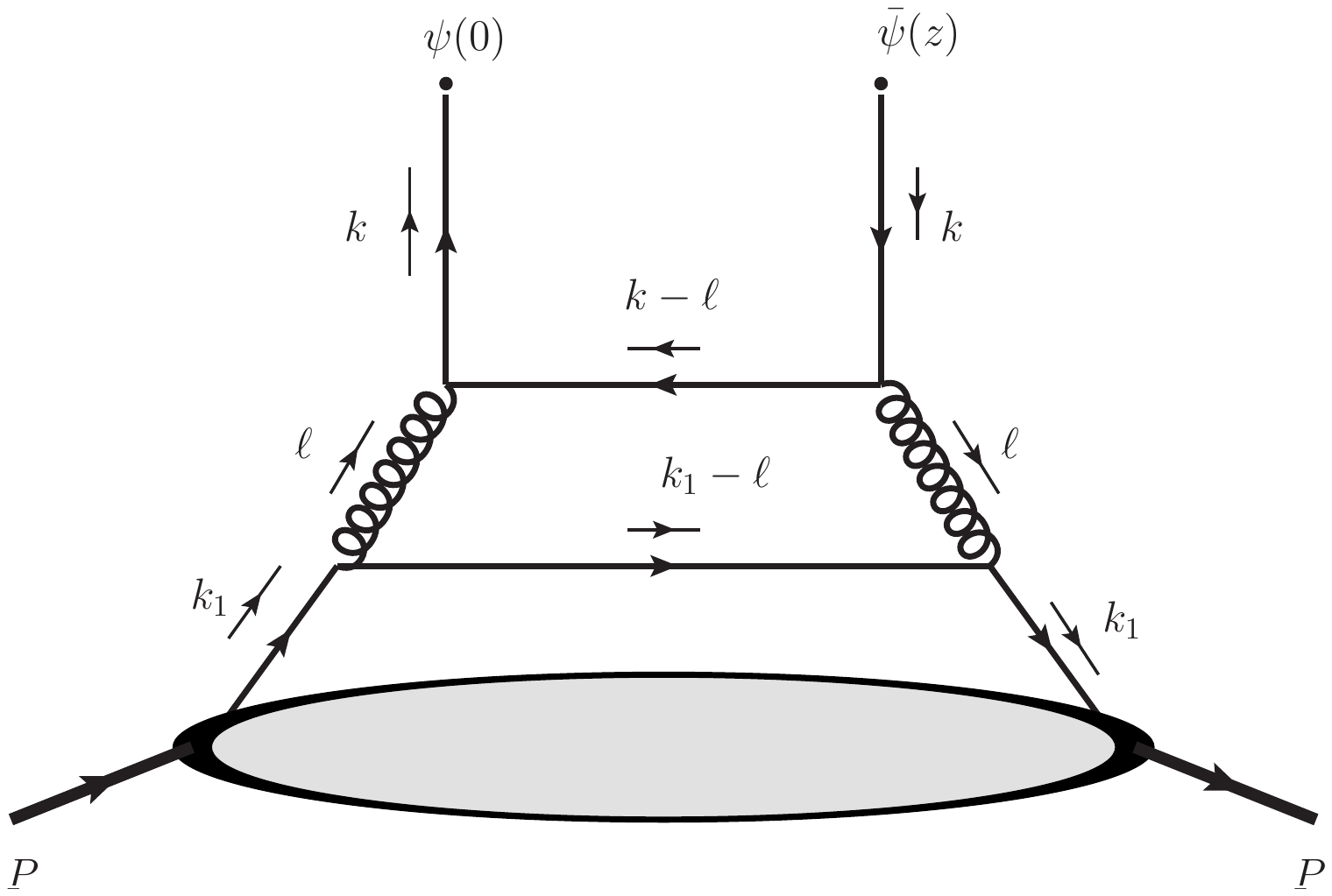}
\vspace{-6cm}
\caption{\label{Fig-S-2} The function $\Phi^{[\gamma^+]}(k)$ at the forth order of strong coupling constant.}
\end{figure}

Let us rewrite Eqn.~(\ref{Phi-1-2}) in the form  of expansion as, see Fig.~\ref{Fig-S-2},
\begin{eqnarray}
\label{Phi-1-2-2}
&&\Phi^{[\gamma^+]}(k) \Rightarrow 
\int (d^4 z) e^{+i kz} \langle P,S | \bar\psi(0) \, 
\gamma^+\, \psi(z) \, \mathbb{S}^{(4)}[\psi,\bar\psi, A] \,| P,S \rangle = 
\nonumber\\
&&
\int (d^4 k_1) (d^4 \ell) 
\text{tr} \big[ 
\gamma^+ S(k) \gamma^\mu S(k-\ell) \gamma^\nu S(k)\big] {\cal D}_{\mu\mu^\prime}(\ell) {\cal D}_{\nu\nu^\prime}(\ell)
\mathcal{F}^{\nu^\prime \mu^\prime}(k_1, \ell),
\end{eqnarray}
where
\begin{eqnarray} 
\label{Phi-1-2-3}
\mathcal{F}^{\nu^\prime \mu^\prime}(k_1, \ell) =  
\int (d^4 \xi) e^{-i k_1 \xi} 
\langle P | \bar\psi(\xi) \gamma^{\nu^\prime} S(k_1-\ell) \gamma^{\mu^\prime} \psi(0) | P\rangle.
\end{eqnarray} 
Focusing on the axial-vector projection of Fierz decomposition of two fermions, 
the function $\mathcal{F}^{\nu^\prime \mu^\prime}(k_1, \ell)$ takes the form of 
\begin{eqnarray} 
\label{Phi-1-2-4}
-4\, \mathcal{F}^{\nu^\prime \mu^\prime}_{(A)}(k_1, \ell) =  
\text{tr}\big[ 
\gamma^{\nu^\prime} S(k_1-\ell) \gamma^{\mu^\prime} \gamma^\alpha \gamma_5
\big]  \Phi^{[\gamma_\alpha \gamma_5]}(k_1)
\end{eqnarray}
with 
\begin{eqnarray}
\label{Phi-A-1}
\Phi^{[\gamma_\alpha \gamma_5]}(k_1)= \int (d^4 \xi) e^{-i k_1 \xi} 
\langle P | \bar\psi(\xi) \gamma_\alpha \gamma_5 \psi(0) | P\rangle.
\end{eqnarray}  
The (sub)structure function $\Phi^{[\gamma_\alpha \gamma_5]}(k_1)$  in Eqn.~(\ref{Phi-A-1}) can be presented in the form of 
${\cal M}$-amplitude written in the momentum representation:
\begin{eqnarray}
\label{Phi-A-1-2}
\delta^{(4)}\big( P_f - P_i \big){\cal M}(k_1) = \delta^{(4)}(0)\Phi^{[\gamma_\alpha \gamma_5]}(k_1) = \langle P| b^+(k_1) b^-(k_1) | P\rangle 
\big[ \bar u(k_1) \gamma_\alpha \gamma_5 u(k_1) \big],
\end{eqnarray} 
where the quark axial-vector combination gives the quark spin, {\it i.e.}
\begin{eqnarray}
\label{S-q}
\big[ \bar u(k_1) \gamma_\alpha \gamma_5 u(k_1) \big] \sim s_\alpha.
\end{eqnarray} 

Hence, the interactions which have been included in the corresponding correlator give the evidences 
for the existence of a new type of parametrizing functions in Eqn.~(\ref{Phi-plus-0}). 
These functions can be treated as the corresponding single quark spin asymmetry inside the unpolarized 
hadron or as the alignment $k_\perp$-dependent functions introduced in \cite{Anikin:2021zxl}. 

To conclude the discussion presented in this subsection, we emphasize that the Lorentz tensor $i \epsilon^{+ - V_\perp s_\perp}$
of Eqn.~(\ref{Phi-plus-0}) contains implicitly the information on the frame system 
where the factorization procedure has been implemented. The separation on the longitudinal and transverse 
components of different Lorentz vectors implies already the certain fixed frame.
Indeed, the corresponding Lorentz invariant would take the form of  $i \epsilon^{n p V s}$ with 
the vectors $p$ and $n$ are being the dominant and sub-sub-dominant 
directions which can be actually chosen in an arbitrary way but in the connection with the kinematics of 
a given process.  As argued in \cite{Anikin:2022ocg}, for the Drell-Yan-like processes 
the most appropriated frame of factorization is given by the Collins-Soper (CS) system where the factorization procedure 
takes the archetypal form. In the CS-frame, the proposed new type of parametrizing functions can be firmly 
singled out. However, the fact of the new function existence might be not easily seen in the other frames. Indeed, 
there is no doubt that any Lorentz invariant (say, an arbitrary scalar 
products) is independent on the chosen frame by construction, but the representation of the given scalar product 
by components depends certainly on the frame.

\section{Conclusions}
\label{Con}

In the presented paper, we have demonstrated that the parametrizing functions which correspond to the 
certain correlator are actually complex functions owing to the discovered role of interactions in the correlator.
As a result of the time-reversal transforms, the $k_\perp$-dependent  
parametrizing functions do not possess the definite properties regarding the replacement $k_\perp \to - k_\perp$.
However, the usual time-reversal properties of functions can be restored after the integration over $k_\perp$.

Also, we have presented an additional evidences for discovering of a new type of parametrizing functions 
which are associated with the inner quark structure defined by the quark spin.

As a practical application of our findings, we suggest that the manifestation of the discussed functional complexity can be observed 
in the Drell-Yan-like (DY-like) processes with the essential role of  the gluon pole contributions \cite{Anikin:2022ocg}.

\acknowledgments

We thank colleagues from the Theoretical Physics Division of NCBJ (Warsaw)
for useful and stimulating discussions.
The work of L.Sz. is supported by the grant 2019/33/B/ST2/02588  
of the National Science Center in Poland.
This work is also supported by the Ulam Program of NAWA No.
PPN/ULM/2020/1/00019.


\appendix
\renewcommand{\theequation}{\Alph{section}.\arabic{equation}}

\section{The compositeness condition for bound states }
\label{App1}

In the hadron physics, the compositeness condition used in the description of bound states
is a key instrument for the practical calculations in the low and intermediate energy regions.  
All details on the compositeness condition and its practical application can be found in the seminal 
works \cite{Efimov:1993zg, Hayashi:1967bjx}. 

In this Appendix, for the convenience of readers, 
we present the main substantial points which should clarify the features of the compositeness condition for 
the high energy and QFT community where it is not widely known.

Let us consider two approaches/models with the interaction Lagrangians defined as 
(here we use the symbolical forms where the corresponding coordinate dependences and 
the tensor structures are neglected)
\begin{eqnarray}
\label{Lag-1}
\mathcal{L}^I_\psi = \lambda_0 (\bar\psi\, \psi)^2, \quad 
\mathcal{L}^I_Y = g_0 M (\bar\psi\, \psi),
\end{eqnarray}  
where the coupling constants and all fields are assumed to be bare (or unphysical) variables; $M$ and $\psi$ stand for the 
boson (hadron) and quark states. 

Notice that the interaction Lagrangian $\mathcal{L}^I_\psi$ resembles formally the Nambu--Jona-Lasinio (NJL) model,
while the interaction Lagrangian $\mathcal{L}^I_Y$ of the Yukawa-type takes the form of the NJL-model after the 
bosonization procedure \cite{Klevansky:1992qe}.
As well-known, the approach with $\mathcal{L}^I_\psi$ in the $D=4$ space 
refers to the unrenormalizable class of theories. In what follows we do not dwell on the questions of renormalization of this 
approach, instead we express all quantities appearing within this approach via the renormalized quantities 
that are derived in the approach with $\mathcal{L}^I_Y$. 

The compositeness condition ensures that the description of the physical processes within both two approaches
with $\mathcal{L}^I_\psi$  and $\mathcal{L}^I_Y$ coincides completely. 
We stress that the hadron states are considered as the bound states of quarks
in these approaches. 
In order to demonstrate that, based on $\mathcal{L}^I_Y$ we first calculate the two-point Green function of bosons within the chain approximation 
(which is fully enough for our discussion). It reads
\begin{eqnarray}
\label{GF-1}
G^{(2)}_Y(s) \equiv D_M(s) = \frac{1}{m_0^2 - s - g_0^2 \Sigma(s)},
\end{eqnarray}
where the mass operator $\Sigma$ is determined by the quark $1$-loop integration which is divergent as a logarithm.
The physical boson with mass $m^2$ corresponds to the pole of this Green function which is a function of $s$:
\begin{eqnarray}
\label{GF-2}
m_0^2 - s - g_0^2 \Sigma(s) \Big|_{s=m^2} = 0.
\end{eqnarray}
Hence, using the Taylor expansion of $\Sigma$ around $s=m^2$, we have 
\begin{eqnarray}
\label{GF-3}
&&D_M(s) = \frac{1}{m^2 - s} \cdot
\frac{1}{ 1 + g^2_0 \Sigma^\prime(m^2) +  g_0^2 \widetilde{\Sigma}(m^2, s)},
\nonumber\\
&&
\Sigma^\prime(m^2) = \frac{\partial \Sigma(s)}{\partial s} \Big |_{s=m^2},
\quad 
\widetilde{\Sigma}(m^2, s)=\frac{\Sigma(m^2, s)}{s-m^2}
\end{eqnarray}
where 
$\Sigma(m^2, s)$ implies the finite residual term of the Taylor expansion.
Eqn.~(\ref{GF-3}) can be rewritten as 
\begin{eqnarray}
\label{GF-4}
D_M(s) = \frac{\mathbb{Z}_M}{m^2 - s} \cdot
\frac{1}{ 1 + g_R^2 \widetilde{\Sigma}(m^2, s)},
\end{eqnarray}
where
\begin{eqnarray}
\label{GF-5}
\mathbb{Z}_M= \frac{1}{1 + g^2_0 \Sigma^\prime(m^2)}\,\,\, \text{and}\,\,\,
g^2_R= \frac{g^2_0}{1 + g^2_0 \Sigma^\prime(m^2)}\equiv \mathbb{Z}_M g^2_0.
\end{eqnarray}
With the help of these equations, one can express the renormalization constant through the renormalized coupling constant as 
\begin{eqnarray}
\label{GF-6}
\mathbb{Z}_M= 1 - g^2_R \Sigma^\prime(m^2).
\end{eqnarray}

Then, we calculate the four-point Green function with the same interaction Lagrangian $\mathcal{L}^I_Y$. We obtain that
\begin{eqnarray}
\label{GF-7}
G^{(4)}_Y(s) \equiv \varGamma(s) = - \frac{g_0^2}{m_0^2 - s - g_0^2 \Sigma(s)} = -
\frac{g_R^2}{m^2 - s} \cdot \frac{1}{1 + g_R^2 \widetilde{\Sigma}(m^2, s)}.
\end{eqnarray}

On the other hand, we can calculate the four-point Green function within the approach with $\mathcal{L}^I_\psi$.
In the chain approximation, we derive that 
\begin{eqnarray}
\label{GF-2-1}
G^{(4)}_\psi (s) = \frac{\lambda_0}{1 + \lambda_0 \Sigma(s)},
\end{eqnarray} 
where the function $\Sigma(s)$ is the same function as in Eqn.~(\ref{GF-1}). Again, this Green function considered 
as the scattering amplitude has a pole which corresponds to the physical boson with $m^2$ provided 
 \begin{eqnarray}
\label{GF-2-2}
1 + \lambda_0 \Sigma(s) \Big|_{s=m^2}=0 \quad \Longrightarrow \quad \lambda_0 = - \frac{1}{\Sigma(m^2)}.
\end{eqnarray}
Inserting the second equation of Eqn.~(\ref{GF-2-2}) into Eqn.~(\ref{GF-2-1}), we get that 
\begin{eqnarray}
\label{GF-2-3}
G^{(4)}_\psi (s) = \frac{1}{\Sigma(s) - \Sigma(m^2)} =
\frac{1}{\Sigma^\prime(m^2) (s-m^2) + \Sigma(m^2, s)}.
\end{eqnarray}
Having multiplied and divided Eqn.~(\ref{GF-2-3}) by $g^2_R$, we obtain 
\begin{eqnarray}
\label{GF-2-4}
G^{(4)}_\psi (s) = - \frac{g^2_R}{m^2 - s} \cdot \frac{1}{ g^2_R \Sigma^\prime(m^2) }\cdot 
\frac{1}{ 1 + \widetilde{\Sigma}(m^2, s) / \Sigma^\prime(m^2) }.
\end{eqnarray} 
In addition, we note that 
\begin{eqnarray}
\label{GF-2-5}
g^2_R \Sigma^\prime(m^2) =1- \mathbb{Z}_M, \quad 
\frac{\widetilde{\Sigma}(m^2, s)}{\Sigma^\prime(m^2)} = \frac{g^2_R \widetilde{\Sigma}(m^2, s)}{1- \mathbb{Z}_M}.
\end{eqnarray} 
Thus, we derive that the four-point Green function within the approach with $\mathcal{L}^I_\psi$ reads 
\begin{eqnarray}
\label{GF-2-6}
G^{(4)}_\psi (s) = - \frac{g^2_R}{m^2 - s} \cdot \frac{1}{ 1- \mathbb{Z}_M }\cdot 
\frac{1}{ 1 + g^2_R \widetilde{\Sigma}(m^2, s)/(1- \mathbb{Z}_M)}.
\end{eqnarray} 
According to the compositeness condition, 
the four-point Green functions (as the scattering amplitudes) of  
Eqns.~(\ref{GF-7}) and (\ref{GF-2-6}) should describe the same physics, {\it i.e.}
\begin{eqnarray}
\label{CompCond-1}
G^{(4)}_Y (s) = G^{(4)}_\psi (s).
\end{eqnarray}
This is possible if and only if we deal with the condition 
\begin{eqnarray}
\label{CompCond-2}
\mathbb{Z}_M=0.
\end{eqnarray}
In other words, the compositeness condition states that the physical hadron state is always dressed by the quark-gluon interaction:
$M(x) = \mathbb{Z}_M^{1/2} M_R(x) = 0$ due to Eqn.~(\ref{CompCond-2}).


\end{document}